\documentclass{l4dc2027}

\usepackage{algorithm}
\usepackage{algpseudocode}
\algdef{SE}[FOR]{MyFor}{EndMyFor}[1]%
  {\algorithmicfor\ #1\ \algorithmicdo}%
  {\algorithmicend\ \algorithmicfor}

\title[Learning Control Lyapunov Functions]{Joint Learning of Control Lyapunov Functions and Controllers via Stable-by-Design Surrogates}
\usepackage{times}
\usepackage{booktabs}
\usepackage{tikz}
\usepackage{pgfplots}
\usepackage{subcaption}
\usepackage{mathtools}
\newtheorem{assumption}{Assumption}
\usepgfplotslibrary{groupplots}
\pgfplotsset{compat=1.16}
\definecolor{cLQR}{RGB}{200,60,60}
\definecolor{cLearned}{RGB}{30,90,200}
\definecolor{cReg}{RGB}{230,140,20}
\definecolor{cSon}{RGB}{30,140,70}
\definecolor{cQPA}{RGB}{140,60,170}
\definecolor{cQPB}{RGB}{110,80,50}
\definecolor{cA1}{RGB}{68,1,84}
\definecolor{cA2}{RGB}{59,82,139}
\definecolor{cA3}{RGB}{0,125,125}
\definecolor{cA4}{RGB}{94,170,60}
\definecolor{cA5}{RGB}{200,150,0}
\newcommand{\phasetrajs}[3]{
  \foreach \i in {0,...,#3}{%
    \pgfmathtruncatemacro{\xc}{2*\i}\pgfmathtruncatemacro{\yc}{2*\i+1}%
    \addplot[#2,forget plot] table[x index=\xc,y index=\yc]{#1};}}
\pgfplotsset{ftrace/.style={width=0.49\linewidth,height=0.40\linewidth,tick label style={font=\scriptsize},label style={font=\scriptsize},title style={font=\scriptsize},unbounded coords=jump},
  ltime/.style={ftrace,xlabel={$t$ [s]},xmin=0,xmax=40}}

\newcommand{\vdptraces}[1]{%
  \addplot[cLQR,dashed,line width=0.7pt]
    table[x=t,y=#1]{figs/vdp_ts_LQR.dat};
  \addplot[cSon,dashed,line width=0.7pt]
    table[x=t,y=#1]{figs/vdp_ts_SontagA.dat};
  \addplot[cSon,line width=0.7pt]
    table[x=t,y=#1]{figs/vdp_ts_SontagB.dat};
  \addplot[cQPA,dashed,line width=0.7pt]
    table[x=t,y=#1]{figs/vdp_ts_QPA.dat};
  \addplot[cQPA,line width=0.7pt]
    table[x=t,y=#1]{figs/vdp_ts_QPB.dat};
  \addplot[cA3,line width=0.7pt]
    table[x=t,y=#1]{figs/vdp_ts_learned_orig0.1.dat};
  \addplot[cReg,line width=0.7pt]
    table[x=t,y=#1]{figs/vdp_ts_learned_orig0.5.dat};
  \addplot[cLearned,line width=0.7pt]
    table[x=t,y=#1]{figs/vdp_ts_learned_sat0.5.dat};
}

\author{%
 \Name{Patricia Pauli} \Email{p.d.pauli@tue.nl}\\
 \addr Mechanical Engineering Department, Eindhoven University of Technology, Eindhoven, The Netherlands
 \AND
 \Name{Matthieu Barreau} \Email{barreau@kth.se}\\
 \addr Division of Decision and Control Systems, KTH Royal Institute of Technology, Stockholm, Sweden%
}

\begin{document}

\maketitle

\begin{abstract}%
Training and verifying neural network-based control Lyapunov functions (CLFs) can be computationally demanding. To address this challenge, we propose a framework that jointly learns a CLF and a stabilizing controller using structured neural networks and a projection-based parameterization. Specifically, we employ Polyak–Łojasiewicz networks (PLNets), which parameterize nonconvex, positive-definite CLF candidates, and construct a stable-by-design surrogate of the closed-loop dynamics that satisfies the Lyapunov decrease condition by construction. The controller combines linear state feedback with a higher-order correction term resembling Sontag's universal formula and is trained to align the resulting closed-loop dynamics with the stable-by-design surrogate. Unlike classical approaches based on Sontag's formula, our framework does not require a known CLF; instead, it learns the CLF and controller simultaneously. We establish local and global stability guarantees for the learned controller and demonstrate the effectiveness of the proposed approach through numerical examples.%
\end{abstract}


\section{Introduction}
Learning-based control has emerged as a promising approach for the design of controllers for complex nonlinear dynamical systems. However, providing stability guarantees for learned controllers remains challenging. Control Lyapunov functions (CLFs) offer a systematic framework for establishing stability and synthesizing stabilizing feedback laws. Given a valid CLF for a control-affine system, a controller can be constructed using Sontag's universal formula \citep{sontag1989universal} or control Lyapunov function quadratic programs (CLF-QPs) \citep{ames2014control}. Both approaches, however, require a suitable CLF to be known a priori, whose construction is generally difficult for nonlinear systems.

To address this challenge, neural networks have been increasingly employed to parameterize and learn Lyapunov functions \citep{richards2018lyapunov,chang2019neural,zhou2022neural,wu2023neural}. Several approaches jointly learn neural Lyapunov functions and control policies, often using counterexample-guided training and formal verification to establish closed-loop stability \citep{chang2019neural,zhou2022neural,yang2024lyapunov}. While these methods provide considerable flexibility, training and verifying the Lyapunov decrease condition can be computationally expensive and may require repeated optimization and verification steps. This motivates the development of learning architectures that incorporate Lyapunov conditions directly into their parameterization.

To learn stable-by-design dynamics, \citet{kolter2019learning} jointly learn a neural dynamics model and a Lyapunov function, projecting the dynamics to satisfy the Lyapunov decrease condition by construction, thereby eliminating the need to enforce the condition during training or verify it afterward. Building on this approach, \citet{harting2026locally} extend the framework to locally exponentially stable neural dynamics whose regions of attraction are exactly characterized by sublevel sets of jointly learned Lyapunov functions. Extending the learning of stable dynamics to control synthesis, \citet{min2023data} jointly learn a dynamics model, a controller, and a Lyapunov function from data, using a projection-based approach to enforce closed-loop stability.

Inspired by these approaches, we develop a framework for jointly learning a CLF and a stabilizing feedback controller for known, potentially unstable nonlinear systems, rather than learning the underlying dynamics. To this end, we introduce a stable-by-design surrogate of the closed-loop dynamics that satisfies the Lyapunov decrease condition by construction. The controller and CLF are jointly learned by minimizing the discrepancy between the actual closed-loop dynamics and this surrogate. The controller combines linear feedback, derived from the linearization of the nonlinear system, with an additional term inspired by Sontag's universal formula, without requiring a valid CLF in advance.

To parameterize the CLF, we employ Polyak–Łojasiewicz networks (PLNets) introduced in \citep{wang2024monotone}. Based on bi-Lipschitz neural networks, PLNets provide expressive, potentially nonconvex, Lyapunov candidates with a unique global minimum. By construction, they satisfy positive definiteness, radial unboundedness, and $V(0)=0$, leaving the Lyapunov decrease condition as the central requirement for stability. Combined with the projection-based surrogate formulation, this enables the joint learning of Lyapunov functions and controllers while exploiting structural properties to simplify training and verification.

The remainder of this paper is organized as follows. Section \ref{sec:problem} presents the problem statement. Section \ref{sec:learning_framework} introduces the proposed learning-based control framework and establishes stability guarantees for the resulting closed-loop system, and Section \ref{sec:experiments} illustrates its application through two nonlinear control examples.



\section{Problem Statement}
\label{sec:problem}

Consider the nonlinear control-affine system
\begin{equation}
    \dot x=f(x)+g(x)u,
    \label{eq:system}
\end{equation}
where $x(t)\in\mathbb R^n$ is the state,
$u(t)\in\mathbb R^m$ is the control input,
$f:\mathbb R^n\to\mathbb R^n$, and
$g:\mathbb R^n\to\mathbb R^{n\times m}$.

\begin{assumption}[System Regularity]
\label{ass:system}
The functions $f$ and $g$ are locally
Lipschitz continuous on $\mathbb R^n$,
with $f(0)=0$. Moreover, $f$ is
differentiable at the origin, and the pair
$(A,B)$, defined by $A\coloneqq \frac{\partial f}{\partial x} (0)$, $B\coloneqq g(0)$, is stabilizable.
\end{assumption}

Our objective is to design a state-feedback controller
$u=\pi(x)$ with $\pi(0)=0$ such that the origin
of the resulting closed-loop system is globally
asymptotically stable. We address this objective using CLFs.

\begin{definition}[Control Lyapunov Function]
A continuously differentiable function
$V:\mathbb R^n\to\mathbb R_{\geq0}$
is a \emph{control Lyapunov function}
if it is positive definite, i.e., $V(0)=0$ and
$V(x)>0$ for all $x\neq 0$, radially unbounded,
and for every $x\neq0$, there exists
$u(t)\in\mathbb R^m$ such that
\begin{equation}
    \dot V(x,u)
    =L_fV(x)+L_gV(x)u<0,
    \label{eq:decrease}
\end{equation}
where
\begin{equation*}
    L_fV(x)=\nabla V(x)^\top f(x),
    \qquad
    L_gV(x)=\nabla V(x)^\top g(x)
\end{equation*}
denote the Lie derivatives of $V$ along
$f$ and $g$, respectively.
\end{definition}

\section{Learning-Based Control Framework}
\label{sec:learning_framework}

We propose a learning-based control framework that combines a locally stabilizing linear controller with a nonlinear augmentation, in the spirit of Sontag's universal formula. We first present our parameterization of CLF candidates by PLNets in Subsection \ref{sec:PD_CLFs} and our chosen controller parameterization consisting of a linear and a higher-order term in Subsection \ref{sec:controller}. Then, we show how the controller and CLF are learned in Subsection \ref{ref:learning}. Afterward, we present local and global stability results for the resulting controller in Subsections \ref{sec:local} and \ref{sec:global}, respectively.

\subsection{CLF Parameterization}
\label{sec:PD_CLFs}

We parameterize the candidate CLF using a PLNet:
\begin{equation}
    V_\theta(x)
    =\|\mathcal G_\theta(x)-\mathcal G_\theta(0)\|^2,
    \label{eq:plnet}
\end{equation}
where $\|\cdot\|$ denotes the Euclidean norm and $\mathcal G_\theta:\mathbb R^n\to\mathbb R^n$ is a bi-Lipschitz network (BiLipNet). The BiLipNet architecture consists of alternating
feed-through networks (FTNs) and orthogonal affine
layers, with trainable parameters collected in
$\theta$ \citep{wang2024monotone}. 
By construction, $\mathcal G_\theta$ satisfies
\begin{equation}
    \mu\|x-y\|
    \leq \|\mathcal G_\theta(x)-\mathcal G_\theta(y)\|
    \leq L\|x-y\|,
    \label{eq:bilipschitz}
\end{equation}
for all $x,y\in\mathbb R^n$, with prescribed
constants $0<\mu\leq L$. Consequently,
\begin{equation*}
    \mu^2\|x\|^2
    \leq V_\theta(x)
    \leq L^2\|x\|^2,
    \label{eq:plnet_bounds}
\end{equation*}
ensuring positive definiteness and radial
unboundedness.

Differentiability and \eqref{eq:bilipschitz}
imply that the Jacobian
$J_{\mathcal G_\theta}(x)
:=\frac{\partial\mathcal G_\theta}{\partial x}(x)$
satisfies
$\sigma_{\min}(J_{\mathcal G_\theta}(x))
\geq\mu$ and
$\|J_{\mathcal G_\theta}(x)\|_2\leq L$,
where $\sigma_{\min}$ denotes the
smallest singular value. Since
\begin{equation*}
    \nabla V_\theta(x)
    =2J_{\mathcal G_\theta}(x)^\top
    \bigl(\mathcal G_\theta(x)
    -\mathcal G_\theta(0)\bigr),
\end{equation*}
we obtain
\begin{equation}
    2\mu^2\|x\|
    \leq\|\nabla V_\theta(x)\|
    \leq 2L^2\|x\|.
    \label{eq:plnet_gradient_bounds}
\end{equation}
Thus, $\nabla V_\theta(0)=0$ and
$\nabla V_\theta(x)\neq0$ for all $x\neq0$.

\subsection{Controller parameterization}\label{sec:controller}
While the PLNet ensures positive definiteness and radial unboundedness, the Lyapunov decrease condition must be addressed through the controller design. Consider the control-affine system \eqref{eq:system}. We first design a feedback gain
$K\in\mathbb{R}^{m\times n}$ such that
$u_{\mathrm{lin}}(x)=-Kx$ stabilizes the linearization of \eqref{eq:system} at the origin. In particular, we choose $K$ and $P\succ0$ satisfying
\begin{equation}
    (A-BK)^\top P+P(A-BK)\prec0.
    \label{eq:linear_lyapunov}
\end{equation}

Let the CLF candidate $V_\theta:\mathbb{R}^n\to\mathbb{R}_{\geq0}$ and a design function $\omega_\phi:\mathbb{R}^n\to\mathbb{R}_{\geq0}$
be PLNets parameterized by $\theta$ and $\phi$, respectively. We consider the feedback law
\begin{equation}
    u_\eta(x)
    =-Kx-
    \frac{L_gV_\theta(x)^\top}
    {\|L_gV_\theta(x)\|^2+\varepsilon}
    [h_\eta(x)]_+,
    \label{eq:learned_controller}
\end{equation}
where $\eta=(\theta,\phi)$,  $[a]_+=\max\{a,0\}$ and $h_\eta(x) \coloneqq L_fV_\theta(x)-L_gV_\theta(x)Kx+\|\nabla V_\theta(x)\|\omega_\phi(x)$, and $\varepsilon>0$. 
The nonlinear augmentation in \eqref{eq:learned_controller}
is activated whenever the nominal linear controller
fails to achieve the Lyapunov decrease
\begin{equation}
    L_fV_\theta(x)-L_gV_\theta(x)Kx \leq - \|\nabla V_\theta(x)\|\omega_\phi(x).
    \label{eq:prescribed_Lyapunov_decrease}
\end{equation}
Here, $\kappa(x)=|\nabla V_\theta(x)|\omega_\phi(x)$ specifies the desired decrease rate. To provide additional flexibility during controller synthesis, we jointly learn $\omega_\phi$ and $V_\theta$ rather than prescribing the decrease rate a priori. We parameterize $\omega_\phi$ as a PLNet with a prescribed positive-definite lower bound, ensuring $\kappa(x)>0$ for all $x\neq 0$. Alternatively, one could choose $\omega_\phi(x)=\alpha V_\theta(x)$ with $\alpha>0$ to reduce the number of trainable parameters. Finally, the regularization parameter $\varepsilon$ in \eqref{eq:learned_controller} ensures that the feedback law remains well-defined as $L_gV_\theta(x)$ approaches zero.



\begin{remark}[Normalized Design Function]
\label{rem:normalization}
To mitigate excessive control effort far from the
origin, we can replace $\omega_\phi(x)$ in \eqref{eq:learned_controller} by
\begin{equation}
    \widetilde{\omega}_\phi(x)
    = \frac{\omega_\phi(x)}{1+\|x\|}.
\end{equation}
This reduces the prescribed decrease rate $\kappa$ for
large $\|x\|$ while preserving positive
definiteness and the local behavior near
the origin. 
\end{remark}

\subsection{Learning using a stable-by-design closed-loop surrogate }\label{ref:learning}

The framework jointly learns a CLF and a controller by minimizing the discrepancy between the closed-loop vector field
\begin{equation}
    f_\eta^{\mathrm{cl}}(x)
    =f(x)+g(x)u_\eta(x)
    \label{eq:learned_closed_loop}
\end{equation}
and a stable-by-design surrogate. Following the projection-based approach of \citet{kolter2019learning,harting2026locally}, we define the surrogate dynamics as
\begin{equation}
    \hat f_\eta^{\mathrm{cl}}(x)
    =f_\eta^{\mathrm{cl}}(x)-q_\theta(x)
    \left[
        q_\theta(x)^\top f_\eta^{\mathrm{cl}}(x)
        +\omega_\phi(x)
    \right]_+,
    \quad
    q_\theta(x)=
    \frac{\nabla V_\theta(x)}
    {\|\nabla V_\theta(x)\|}
    \label{eq:surrogate_closed_loop}
\end{equation}
for $x\neq 0$, with $\hat f_\eta^{\mathrm{cl}}(0)=0$. The following proposition establishes global asymptotic stability of the surrogate dynamics.

\begin{proposition}
Suppose Assumption~\ref{ass:system} holds. Then, for every $\varepsilon>0$, the origin
of the surrogate system
$\dot x=\hat f_\eta^{\mathrm{cl}}(x)$
is globally asymptotically stable.
\end{proposition}

\begin{proof}
By Assumptions~\ref{ass:system}
and \eqref{eq:plnet_gradient_bounds}, $f$, $g$, and
$\nabla V_\theta$ are continuous.
Together with the continuity of $\omega_\phi$
and $\varepsilon>0$, this ensures that
$u_\eta$ and $f_\eta^{\mathrm{cl}}$ are
continuous. Moreover,
$\nabla V_\theta(0)=0$ implies
$u_\eta(0)=0$, and hence
$f_\eta^{\mathrm{cl}}(0)=0$.

Since $\nabla V_\theta(x)\neq0$ for
$x\neq0$, the surrogate vector field \eqref{eq:surrogate_closed_loop}
is continuous away from the origin.
At the origin, using $\|q_\theta(x)\|=1$,
we obtain
\begin{equation*}
    \|\hat f_\eta^{\mathrm{cl}}(x)\|
    \leq 2\|f_\eta^{\mathrm{cl}}(x)\|
    +|\omega_\phi(x)|
    \to0
    \quad\text{as }x\to0.
\end{equation*}
Thus, $\hat f_\eta^{\mathrm{cl}}$ is
continuous on $\mathbb R^n$.

For all $x\neq0$, the derivative of
$V_\theta$ along the surrogate dynamics
satisfies
\begin{align}
    \dot V_\theta(x)
    &=\|\nabla V_\theta(x)\|
    \min\left\{
        q_\theta(x)^\top f_\eta^{\mathrm{cl}}(x),
        -\omega_\phi(x)
    \right\} \leq-\|\nabla V_\theta(x)\|
    \omega_\phi(x)<0,
    \label{eq:surrogate_lyapunov_decrease}
\end{align}
where strict negativity follows from
\eqref{eq:plnet_gradient_bounds} and
$\omega_\phi(x)>0$ for $x\neq0$.

Since $V_\theta$ is positive definite and
radially unbounded, as it is parameterized by a PLNet, its sublevel sets are
compact and, by \eqref{eq:surrogate_lyapunov_decrease},
forward invariant. Continuity of $\hat f_\eta^{\mathrm{cl}}$ ensures existence of solutions,
while compactness and forward invariance
guarantee their forward completeness.
Consequently, global asymptotic stability
of all solutions follows by the standard
Lyapunov direct-method argument; see also
\cite[Theorem~4.2]{khalil2002nonlinear}.
\end{proof}

The projection in \eqref{eq:surrogate_closed_loop} preserves the original closed-loop dynamics whenever $q_\theta(x)^\top f_\eta^{\mathrm{cl}}(x)\leq-\omega_\phi(x)$ and otherwise modifies them to enforce the prescribed Lyapunov decrease condition \eqref{eq:prescribed_Lyapunov_decrease}.

We jointly optimize $\eta=(\theta,\phi)$
to minimize the discrepancy between the
actual and surrogate closed-loop dynamics,
defined as
\begin{equation}
    e_\eta(x):=
    \left\|
        f_\eta^{\mathrm{cl}}(x)
        -\hat f_\eta^{\mathrm{cl}}(x)
    \right\|
    =
    \left[
        q_\theta(x)^\top f_\eta^{\mathrm{cl}}(x)
        +\omega_\phi(x)
    \right]_+,
    \qquad x\neq0.
    \label{eq:surrogate_discrepancy}
\end{equation}
The learning objective over a training domain
$\Omega\subset\mathbb R^n$ is
\begin{equation*}
    \eta^\star\in
    \operatorname*{arg\,min}_\eta
    \int_\Omega e_\eta(x)^2\,\mathrm{d}x.
\end{equation*}
This objective penalizes violations of the
prescribed Lyapunov decrease condition,
enabling joint learning of the candidate
CLF and controller without requiring a
certified CLF beforehand. The learning
procedure is summarized in
Algorithm~\ref{alg:learning_stabilizing_controller}.
The following subsections establish local
exponential stability and a sufficient
condition on the surrogate discrepancy $e_\eta$
for global asymptotic stability of the
learned closed-loop system.


\begin{algorithm}[t]
\caption{Learning a Stabilizing Controller}
\label{alg:learning_stabilizing_controller}
\begin{algorithmic}[1]
    \Require Dynamics $f,g$, domain $\Omega$,
    regularization $\varepsilon>0$, scaling factor $\alpha>0$, iterations $N$, batch size $M$
    \Ensure Learned controller $u_{\eta^\star}$

    \State Linearize the system at the origin
    to obtain $(A,B)$
    \State Compute $K$ and $P\succ0$ satisfying
    \eqref{eq:linear_lyapunov}
    \State Initialize $\eta_0=(\theta_0,\phi_0)$ such that
    $V_{\theta_0}(x)=x^\top Px$ and
    $\omega_{\phi_0}(x)=\alpha V_{\theta_0}(x)$,
    with $\alpha>0$

    \MyFor{$k=1,\ldots,N$}
        \State Sample $\{x_i\}_{i=1}^M\subset\Omega$
        \State Compute the empirical loss
        \Statex \hspace{\algorithmicindent}
        $\displaystyle
        \mathcal L(\eta)=\frac{1}{M}\sum_{i=1}^M
        \left[
            q_\theta(x_i)^\top
            f_\eta^{\mathrm{cl}}(x_i)
            +\omega_\phi(x_i)
        \right]_+^2$
        \State Update $\eta_{k+1}
        =\eta_k-\gamma_k\nabla_\eta\mathcal L_k(\eta_k)$
    \EndMyFor

    \State Set $\eta^\star=\eta_{N+1}$
    \State \Return $u_{\eta^\star}$,
    $V_{\theta^\star}$, $\omega_{\phi^\star}$
\end{algorithmic}
\end{algorithm}

\subsection{Local Exponential Stability}\label{sec:local}

We first establish local stability by showing
that the nonlinear augmentation preserves the
stabilizing behavior of the linear controller
near the origin.

\begin{theorem}[Local Exponential Stability]
\label{thm:local_stability}
Suppose Assumption~\ref{ass:system} holds and let the linear gain $K$ in \eqref{eq:learned_controller} be such that $A-BK$ is Hurwitz. Then, for every fixed
$\varepsilon>0$, the origin of the
closed-loop system \eqref{eq:learned_closed_loop}
is locally exponentially stable.
\end{theorem}

\begin{proof}
By \eqref{eq:plnet_gradient_bounds},
$\nabla V_\theta(x)=O(\|x\|)$ as $x\to0$.
Together with $f(0)=0$ and local Lipschitz
continuity of $f$ and $g$, cmp. Assumption~\ref{ass:system}, this implies
\begin{equation*}
    L_fV_\theta(x)=O(\|x\|^2),
    \qquad
    L_gV_\theta(x)=O(\|x\|).
\end{equation*}
Since $\omega_\phi$ is continuous and
thus locally bounded, $h_\eta(x)=O(\|x\|)$.
For every fixed $\varepsilon>0$, the nonlinear
correction in the controller therefore satisfies
\begin{equation*}
    \frac{L_gV_\theta(x)^\top[h_\eta(x)]_+}
    {\|L_gV_\theta(x)\|^2+\varepsilon}
    =O(\|x\|^2).
\end{equation*}
Consequently, $u_\eta(x)=-Kx+O(\|x\|^2)$.
Since $f$ is differentiable at the origin
and $g$ is locally Lipschitz continuous according to Assumption~\ref{ass:system},
we have $f(x)=Ax+o(\|x\|)$ and
$g(x)=B+O(\|x\|)$. Thus,
\begin{equation*}
    f_\eta^{\mathrm{cl}}(x)
    =(A-BK)x+\rho(x),
    \qquad \rho(x)=o(\|x\|).
\end{equation*}

Moreover, continuity of $f$, $g$,
$\nabla V_\theta$, and $\omega_\phi$,
together with $\varepsilon>0$, ensures
that $u_\eta$ and $f_\eta^{\mathrm{cl}}$
are continuous, so local solutions exist.

Since $A-BK$ is Hurwitz, there exists
$P\succ0$ satisfying
\begin{equation}
    (A-BK)^\top P+P(A-BK)=-I.
    \label{eq:Ricatti_LQR}
\end{equation}
For $W(x)=x^\top Px$, we obtain
$\dot W(x)=-\|x\|^2+2x^\top P\rho(x)$.
Since $\rho(x)=o(\|x\|)$, there exists
$r>0$ such that
$\dot W(x)\leq-\frac12\|x\|^2$
for all $x\in B_r$, where
$B_r=\{x\in\mathbb R^n:\|x\|<r\}$.
Using
$\lambda_{\min}(P)\|x\|^2
\leq W(x)\leq\lambda_{\max}(P)\|x\|^2$,
we conclude that
\begin{equation}
    \dot W(x)
    \leq-\frac{1}{2\lambda_{\max}(P)}W(x),
    \qquad \forall x\in B_r.
    \label{eq:exp_stab}
\end{equation}
A sufficiently small sublevel set of $W$
contained in $B_r$ is forward invariant.
Integrating \eqref{eq:exp_stab} yields $W(x(t))\leq W(x(0)) e^{-t/(2\lambda_{\max}(P))}$, which, together with the quadratic bounds on $W$, establishes local exponential
stability of all closed-loop solutions.

\end{proof}

\begin{remark}[Small Control Property]
The small control property \citep{sontag1989universal}
ensures continuity at the origin of classical
CLF-based feedback laws, such as Sontag's
universal formula. It requires that, for every
$\varepsilon_u>0$, there exists $\delta_u>0$
such that, for every $0<\|x\|<\delta_u$,
there exists a control input
$\|u\|<\varepsilon_u$ satisfying the strict
decrease condition \eqref{eq:decrease}.
Since we assume the linearized system to be stabilizable at the origin and we designed our controller to satisfy
$u_\eta(x)=-Kx+O(\|x\|^2)$, it is therefore continuous at the origin and, by construction, follows the small control property.
\end{remark}


\subsection{Global Asymptotic Stability}\label{sec:global}

We now establish a sufficient condition for global
asymptotic stability based on the discrepancy \eqref {eq:surrogate_discrepancy} between
the learned closed-loop dynamics and the stable-by-design
surrogate.

\begin{theorem}[Global Asymptotic Stability]
\label{thm:global_stability}
Suppose Assumption~\ref{ass:system} holds and let the linear gain $K$ in \eqref{eq:learned_controller} be such that $A-BK$ is Hurwitz.  
If there exists $r>0$ such that the local
exponential stability guarantee of
Theorem~\ref{thm:local_stability} holds
for all initial conditions in $B_r$ and
\begin{equation}
    e_\eta(x)<\omega_\phi(x),
    \qquad \forall x\notin B_r,
    \label{eq:global_discrepancy_condition}
\end{equation}
then the origin of the closed-loop system
\eqref{eq:learned_closed_loop} is globally
asymptotically stable and locally
exponentially stable.
\end{theorem}

\begin{proof}
By Theorem~\ref{thm:local_stability},
the origin is locally exponentially stable,
and its proof establishes continuity of
the closed-loop vector field
$f_\eta^{\mathrm{cl}}$.

For all $x\notin B_r$, the surrogate Lyapunov
decrease condition
\eqref{eq:surrogate_lyapunov_decrease}
and the Cauchy--Schwarz inequality yield
\begin{align}
    \dot V_\theta(x)=\nabla V_\theta(x)^\top
    \hat f_\eta^{\mathrm{cl}}(x)
    +\nabla V_\theta(x)^\top
    \left(f_\eta^{\mathrm{cl}}(x)
    -\hat f_\eta^{\mathrm{cl}}(x)\right)
    \leq-\|\nabla V_\theta(x)\|
    \left(\omega_\phi(x)-e_\eta(x)\right)<0,
    \label{eq:strict_decrease_global}
\end{align}
where strict negativity follows from
\eqref{eq:plnet_gradient_bounds} and
\eqref{eq:global_discrepancy_condition}.

For any initial condition $x(0)\in\mathbb R^n$,
choose
\begin{equation*}
    c>\max\left\{
        V_\theta(x(0)),
        \max_{x\in\overline B_r}V_\theta(x)
    \right\},
\end{equation*}
where $\overline B_r=\{x:\|x\|\leq r\}$,
and define
$\Omega_c=\{x:V_\theta(x)\leq c\}$.
Since $\overline B_r\subset
\operatorname{int}(\Omega_c)$,
the boundary $\partial\Omega_c$
lies outside $B_r$. By
\eqref{eq:strict_decrease_global},
$\dot V_\theta<0$ on $\partial\Omega_c$,
so $\Omega_c$ is forward invariant.
Moreover, $\Omega_c$ is compact by
continuity and radial unboundedness of
$V_\theta$. Together with continuity of
$f_\eta^{\mathrm{cl}}$, this ensures
existence and forward completeness of 
all solutions starting in $\Omega_c$.

Now consider the compact set
$\mathcal S_c=\Omega_c\setminus B_r$.
If $\mathcal S_c$ is nonempty, continuity and strict
negativity of $\dot V_\theta$ on $\mathcal S_c$
imply that there exists $\delta_c>0$
such that $\dot V_\theta(x)\leq-\delta_c$
for all $x\in \mathcal S_c$.
Suppose a solution never enters $B_r$.
By forward invariance of $\Omega_c$,
it remains in $\mathcal S_c$ for all $t\geq0$,
implying $V_\theta(x(t))
    \leq V_\theta(x(0))-\delta_c t$. This contradicts $V_\theta\geq0$ for
sufficiently large $t$. Hence, every
solution enters $B_r$ in finite time
and subsequently converges to the origin
by Theorem~\ref{thm:local_stability}.
Together with local exponential stability,
this establishes global asymptotic
stability of all closed-loop solutions.
\end{proof}

Theorem~\ref{thm:global_stability} provides
a sufficient condition for global asymptotic
stability of the learned closed-loop system.
The discrepancy condition
\eqref{eq:global_discrepancy_condition}
must therefore be verified after training
to certify this guarantee.


\section{Numerical Experiments}
\label{sec:experiments}
We evaluate the proposed framework on two nonlinear control-affine systems to assess its stabilization performance, generalization beyond the training domain, and control effort relative to established CLF-based controllers. We further investigate how the prescribed decrease rate influences the trade-off between convergence speed and control effort.

\paragraph{Systems.}
We consider two
control-affine systems:
\begin{itemize}
    \item \emph{Academic example}
    ($\Omega=[-2,2]^2$):
    $\dot x_1=x_2$,
    $\dot x_2=x_1+x_1^3-0.2x_2+u$.
    The destabilizing cubic nonlinearity
    limits the region of attraction of
    linear feedback.

    \item \emph{Van der Pol oscillator}
    ($\Omega=[-7,7]^2$):
    $\dot x_1=x_2$,
    $\dot x_2=-x_1+(1-x_1^2)x_2+u$,
    with an unstable equilibrium at the origin.
\end{itemize}

\paragraph{Baselines.}
We compare against a linear quadratic regulator (LQR) and Sontag's universal
formula
\begin{equation*}
    u_{\mathrm S}(x)=
    -\frac{a+\sqrt{a^2+\|b\|^4}}{\|b\|^2}b,
    \quad
    a=L_fV(x),\quad b=L_gV(x)^\top,
\end{equation*}
with $u_{\mathrm S}=0$ for $b=0$ \citep{sontag1989universal}, and a
CLF-QP minimizing $\|u\|^2$ subject to
$L_fV+L_gV\,u\leq-\lambda V$.
Both CLF-based controllers use either the quadratic function $V(x)=x^\top Xx$,
where $X$ solves the algebraic Riccati equation of the LQR problem with gain $K=B^\top X$
(the same $K$ as in \eqref{eq:learned_controller}),
or the learned $V_\theta$.

For both systems, $u=-\varphi(x)-Kx$ cancels the respective nonlinearity $\varphi(x)$ and yields
$\dot V=-x^\top(I+K^\top K)x<0$ for $x\neq0$.
Thus, $V=x^\top Xx$ is a global CLF. 
For the learned $V_\theta$, a numerical search over $\Omega$ and $1.4\,\Omega$
(random points projected onto $L_gV_\theta=0$, for all trained networks)
found no violation of the CLF condition, although this is not a formal certificate.
The CLF-QP decay rates $\lambda=1.65$ (academic example) and $\lambda=0.54$ (Van der Pol) were selected empirically on the evaluation initial conditions.

\paragraph{Implementation.}
For both systems, $K$ is the LQR gain
with $Q=R=I$, and $P\succ0$ solves \eqref{eq:Ricatti_LQR}. Both $V_\theta$ and $\omega_\phi$ are
parameterized as PLNets
\citep{wang2024monotone}, with two
feed-through residual layers interleaved
with orthogonal layers. Each residual
layer contains two hidden $\tanh$ layers
of width 16, with Lipschitz bounds
$(\mu,\nu)=(0.5,2)$. 
The networks are initialized as
$V_\theta(x)=x^\top Px$ and
$\omega_\phi(x)=\alpha x^\top Px$,
where $\alpha>0$ scales the underlying BiLipNet
by $\sqrt{\alpha}$ and hence its Lipschitz bounds.
The PLNet structures preserve the prescribed
quadratic bounds throughout training, including
$\omega_\phi(x)\geq
\alpha\lambda_{\min}(P)\|x\|^2$.


Training uses Adam with learning rate
$3\cdot10^{-3}$, cosine decay, and
12\,000 steps with batches of 4096
states sampled uniformly from $\Omega$.
We use five random seeds,
$\varepsilon=0.1$, and $\alpha=1$
for the academic example.
For the Van der Pol oscillator, we compare the original
decrease rate with the normalized
$\omega_\phi(x)/(1+\|x\|)$ from
Remark~\ref{rem:normalization}. 
Although the discrepancy condition \eqref{eq:global_discrepancy_condition} is not formally verified, a grid-based evaluation of the Lyapunov decrease condition over $\Omega$ revealed violations only near the origin.

\paragraph{Evaluation.}
We simulate each controller using the
classical fourth-order Runge--Kutta
method with a step size of $0.005$\,s
over $T=40$\,s, starting from 12 initial conditions inside and 12 outside $\Omega=[-r,r]^2$,
placed at equally spaced angles on the circles of radius $0.9\,r$ and $1.4\,r$, respectively,
and identical for all seeds. 
A trajectory is considered converged if
$\|x(T)\|<10^{-2}$.

We report convergence success, settling
time, input energy $\int_0^T\|u\|^2\,dt$,
and peak input magnitude.
The latter three metrics are evaluated
for converged trajectories starting
inside $\Omega$, reporting medians for
settling time and energy and the maximum
peak input. 

\subsection{Stabilization Performance}
\label{sec:performance}

\begin{table}[t]
\centering
\caption{
Stabilization performance (five seeds). ``In $\Omega$'' and ``beyond'' give the number of converged runs out of 12 (mean over seeds, with $\pm$ the standard deviation if nonzero).
Settling time $t_{\mathrm s}=\inf\{t\in[0,T]:\|x(\tau)\|<10^{-2}\ \forall\tau\in[t,T]\}$, energy and peak input use converged runs from the starts inside $\Omega$, pooled over seeds (medians for $t_{\mathrm s}$ and energy, maximum for the peak).
On the Van der Pol oscillator, Sontag's formula and the CLF-QP with a learned CLF use the $V_\theta$ of the learned $\omega_\phi/(1+\|x\|)$ model with $\alpha=0.5$.
}
\label{tab:results}
\small
\begin{tabular}{lccccc}
    \toprule
    & in $\Omega$ & beyond & $t_{\mathrm s}$ [s] & $\int\!\|u\|^2$ & peak $\|u\|$ \\
    \midrule
\multicolumn{6}{l}{\textit{Academic example}} \\
    LQR & $4$ & $4$ & $5.0$ & $3.5$ & $4$ \\
    Sontag (analytic CLF) & $12$ & $12$ & $6.4$ & $19.3$ & $19.6$ \\
    Sontag (learned CLF) & $12$ & $12$ & $6.2$ & $43.3$ & $45.2$ \\
    CLF-QP (analytic CLF) & $12$ & $12$ & $7.1$ & $18.8$ & $9.87$ \\
    CLF-QP (learned CLF) & $12$ & $12$ & $6.6$ & $16.7$ & $12$ \\
    Learned, $\alpha=1$ (ours) & $12$ & $12$ & $7.3$ & $16.9$ & $18.1$ \\
\midrule
\multicolumn{6}{l}{\textit{Van der Pol oscillator}} \\
    LQR & $12$ & $10$ & $14.8$ & $15.4$ & $16.9$ \\
    Sontag (analytic CLF) & $12$ & $12$ & $18.1$ & $12.1$ & $41.8$ \\
    Sontag (learned CLF) & $12$ & $12$ & $7.9$ & $396$ & $167$ \\
    CLF-QP (analytic CLF) & $12$ & $12$ & $28.8$ & $4.9\cdot10^{4}$ & $7113$ \\
    CLF-QP (learned CLF) & $12$ & $12$ & $16.4$ & $672$ & $1.3\cdot10^{4}$ \\
    Learned $\omega_\phi$, $\alpha=0.5$ & $12$ & $12$ & $6.0$ & $2.0\cdot10^{4}$ & $8614$ \\
    Learned $\omega_\phi$, $\alpha=0.1$ & $12$ & $12$ & $12.1$ & $254$ & $45.4$ \\
    Learned $\omega_\phi/(1+\|x\|)$, $\alpha=0.5$ & $12$ & $12$ & $12.6$ & $166$ & $28.7$ \\
    \bottomrule
\end{tabular}
\end{table}

\paragraph{Academic example.}

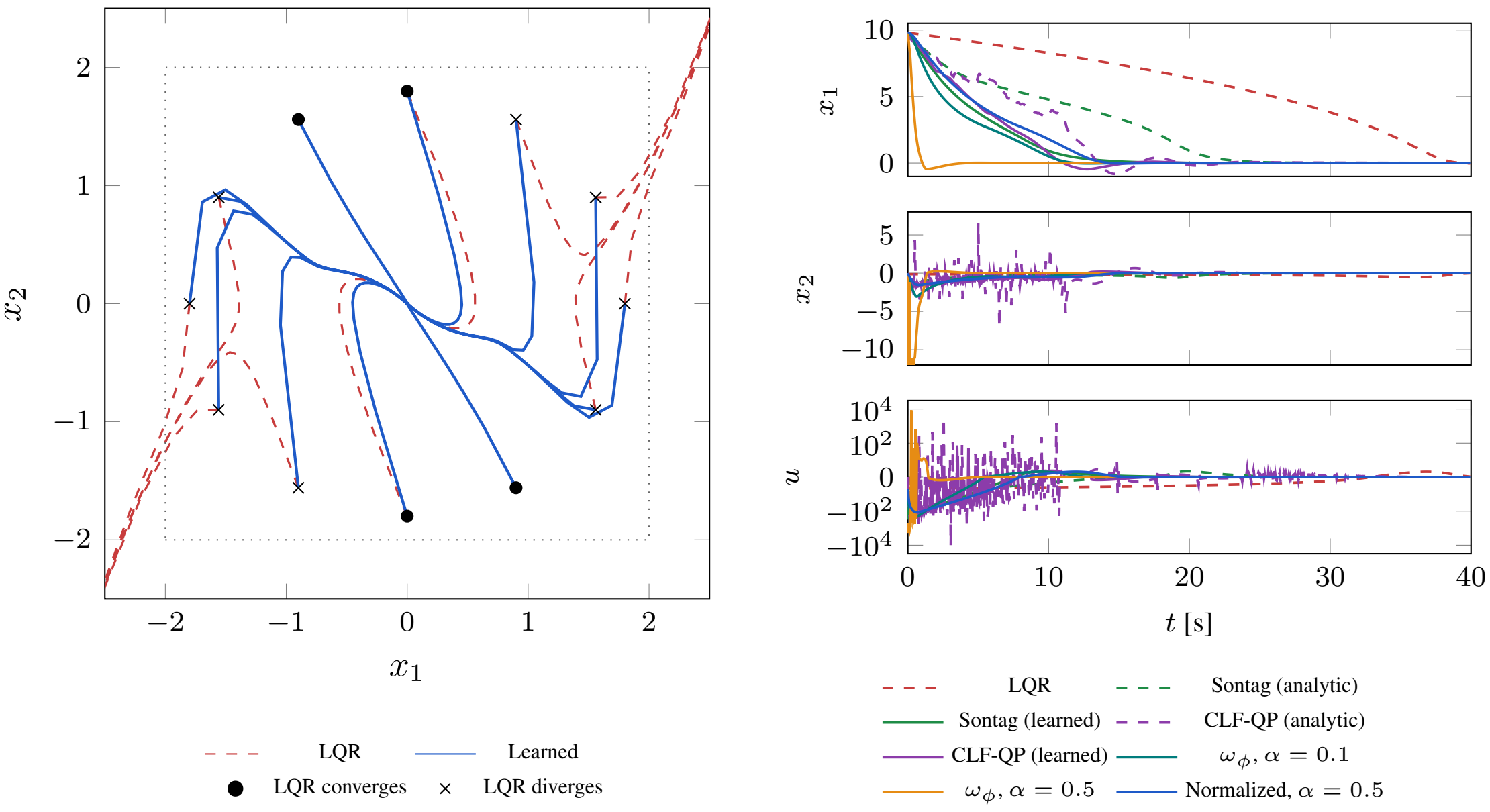
\begin{figure}[t]
\centering

\begin{minipage}[t]{0.46\linewidth}
\vspace{0pt}
\centering
\begin{tikzpicture}
\begin{axis}[
    width=0.85\linewidth,
    height=0.83\linewidth,
    scale only axis,
    xlabel={$x_1$},
    ylabel={$x_2$},
    xmin=-2.5, xmax=2.5,
    ymin=-2.5, ymax=2.5,
    tick label style={font=\scriptsize},
    label style={font=\small},
    legend style={
        font=\tiny,
        at={(0.5,-0.22)},
        anchor=north,
        legend columns=2,
        draw=none
    }
]
    \phasetrajs{figs/academic_lqr_phase.dat}
        {cLQR,dashed,line width=0.6pt}{11}
    \phasetrajs{figs/academic_learned_phase.dat}
        {cLearned,line width=0.8pt}{11}

    \addplot[black!60,dotted,forget plot]
        coordinates {
            (-2,-2) (2,-2) (2,2)
            (-2,2) (-2,-2)
        };

    \addplot[
        only marks,mark=*,
        mark size=1.6pt,black,forget plot
    ] table[x=x,y=y]{figs/academic_starts_ok.dat};

    \addplot[
        only marks,mark=x,
        mark size=2.2pt,black,forget plot
    ] table[x=x,y=y]{figs/academic_starts_fail.dat};

    \addlegendimage{cLQR,dashed}
    \addlegendentry{LQR}
    \addlegendimage{cLearned}
    \addlegendentry{Learned}
    \addlegendimage{only marks,mark=*,black}
    \addlegendentry{LQR converges}
    \addlegendimage{only marks,mark=x,black}
    \addlegendentry{LQR diverges}
\end{axis}
\end{tikzpicture}

\par\smallskip
{\small (a) Academic example.}
\end{minipage}%
\hfill
\begin{minipage}[t]{0.52\linewidth}
\vspace{0pt}
\centering

\pgfplotsset{
    ts/.style={
        width=0.90\linewidth,
        height=0.39\linewidth,
        tick label style={font=\scriptsize},
        label style={font=\scriptsize},
        unbounded coords=jump,
        xmin=0,xmax=40
    }
}

\begin{tikzpicture}
\begin{groupplot}[
    group style={
        group size=1 by 3,
        vertical sep=0.35cm,
        xticklabels at=edge bottom
    },
    ts
]
\nextgroupplot[
    ylabel={$x_1$},
    ymin=-1,ymax=10.5,
    legend to name=leg:ts,
    legend columns=2,
    legend style={font=\tiny,draw=none}
]
    \vdptraces{x1}

    \addlegendimage{cLQR,dashed}
    \addlegendentry{LQR}
    \addlegendimage{cSon,dashed}
    \addlegendentry{Sontag (analytic)}
    \addlegendimage{cSon}
    \addlegendentry{Sontag (learned)}
    \addlegendimage{cQPA,dashed}
    \addlegendentry{CLF-QP (analytic)}
    \addlegendimage{cQPA}
    \addlegendentry{CLF-QP (learned)}
    \addlegendimage{cA3}
    \addlegendentry{$\omega_\phi$, $\alpha=0.1$}
    \addlegendimage{cReg}
    \addlegendentry{$\omega_\phi$, $\alpha=0.5$}
    \addlegendimage{cLearned}
    \addlegendentry{Normalized, $\alpha=0.5$}

\nextgroupplot[
    ylabel={$x_2$},
    ymin=-12,ymax=8
]
    \vdptraces{x2}

\nextgroupplot[
    ylabel={$u$},
    xlabel={$t$ [s]},
    ymin=-4.5,ymax=4.5,
    ytick={-4,-2,0,2,4},
    yticklabels={
        $-10^4$,$-10^2$,$0$,$10^2$,$10^4$
    }
]
    \vdptraces{u}

\end{groupplot}
\end{tikzpicture}

\par\smallskip
\pgfplotslegendfromname{leg:ts}

\par\smallskip
{\small (b) Van der Pol oscillator.}
\end{minipage}

\caption{Closed-loop responses (seed~0).
(a) Academic example: LQR diverges from
8 of 12 initial conditions, whereas the
learned controller converges from all.
(b) Van der Pol oscillator: states and
control input from $x(0)=(9.8,0)$.
The input uses the signed logarithmic scale
$\operatorname{sign}(u)\log_{10}(1+|u|)$;
dense bands indicate rapid input oscillations.}
\label{fig:numerical_comparison}
\end{figure}

Figure~\ref{fig:numerical_comparison}(a) illustrates
the improved performance over LQR, which converges
from only 4 of 12 initial conditions.
The learned controller and all CLF-based baselines
converge from every tested initial condition,
including those outside $\Omega$, across all five
seeds (Table~\ref{tab:results}).
The learned controller achieves comparable settling
times and control energies. Unlike these baselines, the proposed framework jointly learns the controller and Lyapunov function instead of constructing the feedback from an a-priori known CLF through Sontag's formula or online optimization.


\paragraph{Van der Pol oscillator.}

Table~\ref{tab:results} shows that the learned controller converges from
all tested initial conditions, including
those outside $\Omega$ where LQR does
not reach the convergence tolerance.
The CLF-based baselines also achieve
high convergence success, but differ
substantially in control effort.
In particular, Sontag's formula with
the quadratic CLF uses little energy
but converges more slowly, whereas the
CLF-QP produces large control inputs.
Using the learned $V_\theta$ changes
these trade-offs and can further
increase control effort.

The prescribed decrease-rate parameter $\alpha$, introduced in the implementation paragraph, provides a means to adjust the trade-off between convergence speed and control effort. With the original $\omega_\phi$ and $\alpha=0.5$, the learned controller achieves the shortest settling time (6.0\,s), but requires substantial control energy ($2.0\cdot10^4$) and a peak input of $8.6\cdot10^3$. Normalizing the decrease rate, cmp. Remark~\ref{rem:normalization}, reduces these values to 166 and 28.7, respectively, while increasing the settling time to 12.6\,s. Alternatively, reducing $\alpha$ in the original formulation also limits control effort, indicating that both approaches can improve the trade-off between convergence speed and control magnitude. Figure~\ref{fig:numerical_comparison}(b) compares the closed-loop responses from a challenging
initial condition beyond $\Omega$, showing that
controllers with high control effort exhibit
large, rapidly varying inputs.

\section{Conclusions}
\label{sec:conclusions}

We presented a learning-based framework for jointly synthesizing a CLF and a stabilizing controller for nonlinear control-affine systems. The framework employs PLNets to parameterize positive-definite CLFs and constructs stable-by-design surrogate dynamics through a projection-based parameterization to guide controller learning. Unlike conventional CLF-based methods, the proposed approach requires neither a prescribed CLF nor online optimization. We established sufficient conditions for local exponential and global asymptotic stability of the learned closed-loop system. Numerical experiments demonstrated improved stabilization compared with linear feedback and competitive performance relative to Sontag's formula and CLF-QP controllers.

Future work will focus on extending the framework to higher-dimensional systems, incorporating input constraints, and formally verifying the learned controllers.




\subsection*{Generative AI Use}


Generative AI tools were used for writing and editing in text and proofs and for coding. The authors reviewed and verified all AI-assisted content and take responsibility for the final content of the paper.

\acks{We acknowledge financial support from COST Action CA24136 through a short-term scientific mission (STSM) grant. This work was partially supported by the Wallenberg AI, Autonomous Systems and Software Program (WASP) funded by the Knut and Alice Wallenberg Foundation.}

\bibliography{l4dc2027-sample}

\begin{thebibliography}{12}
\providecommand{\natexlab}[1]{#1}
\providecommand{\url}[1]{\texttt{#1}}
\expandafter\ifx\csname urlstyle\endcsname\relax
  \providecommand{\doi}[1]{doi: #1}\else
  \providecommand{\doi}{doi: \begingroup \urlstyle{rm}\Url}\fi

\bibitem[Ames et~al.(2014)Ames, Grizzle, and Tabuada]{ames2014control}
Aaron~D Ames, Jessy~W Grizzle, and Paulo Tabuada.
\newblock Control barrier function based quadratic programs with application to
  adaptive cruise control.
\newblock In \emph{53rd IEEE conference on decision and control}, pages
  6271--6278. IEEE, 2014.

\bibitem[Chang et~al.(2019)Chang, Roohi, and Gao]{chang2019neural}
Ya-Chien Chang, Nima Roohi, and Sicun Gao.
\newblock Neural {L}yapunov control.
\newblock \emph{Advances in Neural Information Processing Systems}, 32, 2019.

\bibitem[Harting et~al.(2026)Harting, Johansson, Tarbouriech, and
  Barreau]{harting2026locally}
Alice Harting, Karl~Henrik Johansson, Sophie Tarbouriech, and Matthieu Barreau.
\newblock Locally stable neural odes with characterized region of attraction.
\newblock \emph{arXiv preprint arXiv:2606.19109}, 2026.

\bibitem[Khalil and Grizzle(2002)]{khalil2002nonlinear}
Hassan~K Khalil and Jessy~W Grizzle.
\newblock \emph{Nonlinear systems}, volume~3.
\newblock Prentice hall, Upper Saddle River, NJ, 2002.

\bibitem[Kolter and Manek(2019)]{kolter2019learning}
J~Zico Kolter and Gaurav Manek.
\newblock Learning stable deep dynamics models.
\newblock \emph{Advances in Neural Information Processing Systems}, 32, 2019.

\bibitem[Min et~al.(2023)Min, Richards, and Azizan]{min2023data}
Youngjae Min, Spencer~M Richards, and Navid Azizan.
\newblock Data-driven control with inherent {L}yapunov stability.
\newblock In \emph{Conference on Decision and Control}, pages 6032--6037. IEEE,
  2023.

\bibitem[Richards et~al.(2018)Richards, Berkenkamp, and
  Krause]{richards2018lyapunov}
Spencer~M Richards, Felix Berkenkamp, and Andreas Krause.
\newblock The {L}yapunov neural network: Adaptive stability certification for
  safe learning of dynamical systems.
\newblock In \emph{Conference on robot learning}, pages 466--476. PMLR, 2018.

\bibitem[Sontag(1989)]{sontag1989universal}
Eduardo~D Sontag.
\newblock A ‘universal’construction of artstein's theorem on nonlinear
  stabilization.
\newblock \emph{Systems \& control letters}, 13\penalty0 (2):\penalty0
  117--123, 1989.

\bibitem[Wang et~al.(2024)Wang, Dvijotham, and Manchester]{wang2024monotone}
Ruigang Wang, Krishnamurthy~Dj Dvijotham, and Ian Manchester.
\newblock Monotone, bi-{L}ipschitz, and polyak-{Ł}ojasiewicz networks.
\newblock In \emph{Proceedings of the 41st International Conference on Machine
  Learning}, volume 235, pages 50379--50399. PMLR, 21--27 Jul 2024.

\bibitem[Wu et~al.(2023)Wu, Clark, Kantaros, and Vorobeychik]{wu2023neural}
Junlin Wu, Andrew Clark, Yiannis Kantaros, and Yevgeniy Vorobeychik.
\newblock Neural lyapunov control for discrete-time systems.
\newblock \emph{Advances in Neural Information Processing Systems},
  36:\penalty0 2939--2955, 2023.

\bibitem[Yang et~al.(2024)Yang, Dai, Shi, Hsieh, Tedrake, and
  Zhang]{yang2024lyapunov}
Lujie Yang, Hongkai Dai, Zhouxing Shi, Cho-Jui Hsieh, Russ Tedrake, and Huan
  Zhang.
\newblock {L}yapunov-stable neural control for state and output feedback: A
  novel formulation.
\newblock In \emph{Proceedings of the 41st International Conference on Machine
  Learning}, volume 235, pages 56033--56046. PMLR, 21--27 Jul 2024.

\bibitem[Zhou et~al.(2022)Zhou, Quartz, De~Sterck, and Liu]{zhou2022neural}
Ruikun Zhou, Thanin Quartz, Hans De~Sterck, and Jun Liu.
\newblock Neural lyapunov control of unknown nonlinear systems with stability
  guarantees.
\newblock \emph{Advances in Neural Information Processing Systems},
  35:\penalty0 29113--29125, 2022.

\end{thebibliography}

\end{document}